\documentclass[12pt]{article}
\usepackage[affil-it]{authblk}

\usepackage{array,multirow,graphicx}
\usepackage{titlesec}
    \titleformat{\section}{\normalfont\Large\filcenter}{\thesection}{1em}{\MakeUppercase}

\usepackage[margin=1.0in]{geometry}
\usepackage{color}
\usepackage{times}
\usepackage{amsmath}
\usepackage{amsfonts}
\usepackage{enumerate}
\usepackage{lscape}
\usepackage{rotating}
\usepackage{epstopdf}

\usepackage{float}
\usepackage[nomarkers]{endfloat}

\def\({\left(}
\def\){\right)}
\def\[{\left[}
\def\]{\right]}

\newcommand{\bmu}{\boldsymbol{\mu}}

\newcommand{\blambda}{\boldsymbol{\lambda}}

\newcommand{\0}{\mathbf{0}}

\newcommand{\blue}{\textcolor{blue}}

\def\be{\begin{eqnarray}}
\def\ee{\end{eqnarray}}
\def\bq{\begin{equation}}
\def\eq{\end{equation}}
\def\bse{\begin{eqnarray*}}
\def\ese{\end{eqnarray*}}

\def\th{\mathrm{th}}

\def\bX{\mathbf{X}}

\def\bY{\mathbf{Y}}
\def\by{\mathbf{y}}

\def\bdSigma{\boldsymbol{\Sigma}}
\def\bdOmega{\boldsymbol{\Omega}}

\begin{document}

\title{A random covariance model for bi-level graphical modeling with application to resting-state fMRI data}
\author{Lin Zhang$^{1\dagger *}$, Andrew DiLernia$^{1\dagger}$, Karina Quevedo$^2$, Jazmin Camchong$^2$, Kelvin Lim$^2$, Wei Pan$^1$}
\affil{
$^1$ Division of Biostatistics, University of Minnesota, Minneapolis, MN, U.S.A. \\
$^2$ Department of Psychiatry, University of Minnesota, Minneapolis, MN, U.S.A. \\
{\normalsize $^\dagger$ Contributed equally to this work. } \\
}
\date{}

\maketitle

\begin{abstract}
This paper considers a novel problem, bi-level graphical modeling, in which multiple individual graphical models can be considered as variants of a common group-level graphical model and inference of both the group- and individual-level graphical models are of interest. Such problem arises from many applications including multi-subject neuroimaging and genomics data analysis. We propose a novel and efficient statistical method, the random covariance model, to learn the group- and individual-level graphical models simultaneously. The proposed method can be nicely interpreted as a random covariance model that mimics the random effects model for mean structures in linear regression. It accounts for similarity between individual graphical models, identifies group-level connections that are shared by individuals in the group, and at the same time infers multiple individual-level networks. Compared to existing multiple graphical modeling methods that only focus on individual-level networks, our model learns the group-level structure underlying the multiple individual networks and enjoys computational efficiency that is particularly attractive for practical use. We further define a measure of degrees-of-freedom for the complexity of the model that can be used for model selection. We demonstrate the asymptotic properties of the method and show its finite-sample performance through simulation studies. Finally, we apply the proposed method to our motivating clinical data, a multi-subject resting-state functional magnetic resonance imaging (fMRI) dataset collected from schizophrenia patients.  

\blue{keywords:} random covariance model; multiple graphical model; bi-level graphical model; graphical lasso; functional connectivity
\end{abstract}


\newpage

\section{Introduction}

The graphical model has been commonly used to depict the conditional dependence among a set of random variables, $\bX=(X_1,\ldots,X_p)$, which is composed of a set of nodes that represent the variables of interest and a number of edges the represent the associations between the nodes it connects. In a Gaussian graphical model (GGM), in which variables follow a multivariate Gaussian distribution, i.e. $\bX \sim \mathcal{N}(\bmu,\bdSigma)$, two nodes/variables are considered to be conditionally independent given all other variables if and only if their corresponding off-diagonal entry in the precision matrix is zero. Thus the problem of learning a graphical model or network, i.e. deciding which nodes are connected by edges, is equivalent to configuring the nonzero structures of the precision matrix, $\bdOmega=\bdSigma^{-1}$. 

Many methods have been developed for GGM by using regularization to induce sparsity in the estimated precision matrix and the corresponding graphical model. Some methods were proposed to identify the locations of non-zero entries in the precision matrix by utilizing a series of regression models, each of which regresses one variable on all others with an $L_1$ penalty on the coefficients (Meinshausen and B\:{u}hlmann, 2006; Cai et al., 2011). These methods do not provide an estimate of the precision matrix itself, but instead configures the nonzero structures of the matrix. Yuan and Lin (2007) proposed another method, the graphical lasso method, which induces a shrunken estimator of the precision matrix by maximizing the penalized log-likelihood with an $L_1$ penalty on the precision matrix. The graphical lasso method can yield estimates for off-diagonal entries in a precision matrix that are exactly zero, and thus renders straightforward edge selection and graphical model inference. A number of efficient algorithms have been developed for computing the estimator as seen in Friedman et al. (2007) and Rothman et al. (2008) among others. The theoretical properties of the graphical lasso method have also been extensively studied by Rothman et al. (2008) and Lam and Fan (2009), who showed the consistency of its estimator in both estimation and model selection. 

Recent studies have considered the problem of joint inference of multiple graphical models. In particular, Guo et al. (2011) proposed a multiple graphical model utilizing a hierarchical penalty that targets the removal of common zeros in the precision matrices. Danaher et al. (2014) developed a joint graphical lasso (JGL) method that introduces similarity between graphs by using an $L_1$ penalty on pairwise differences of the precision matrices. Villa-Vialaneix et al. (2014) proposed a method that shrinks individual correlations to a fixed consensus value. Peterson et al. (2015) provided a Bayesian approach which uses a Markov random field prior to encourage common structures for multiple network inference. Cai et al. (2016) and Tao et al. (2016) proposed methods that estimate multiple precision matrices with the same sparsity structure but different correlations. These methods simultaneously estimate multiple sparse precision matrices and their corresponding graphical models under the assumption that these precision matrices are similar to or the same as each other. 

In this paper we consider a related but different issue, bi-level graphical modeling, in which multiple graphical models can be viewed as variants of a common group-level graphical model and inference of both group- and individual- level networks are of interest. Such a problem arises in many applications including the functional connectivity analysis of neuroimaging data and gene regulatory network analysis of single-cell sequencing data that are collected from multiple subjects or conditions. Current analyses of these data have primarily focused on single subject/condition analysis. However, this type of analyses is limited by the reliability of relevant technologies, does not borrow strength from the data of other subjects/conditions that share the same clinical characteristics, and fails to provide group-level information that might shed light on diagnostic or treatment strategies specific to clinical disorders. Instead, simultaneous inference of both group- and individual-level graphical models is desired in these scenarios, which allows researchers to examine the shared patterns for the subjects/conditions with the same clinical characteristics as well as identify subject/condition-specific alterations for precision medicine. This problem of bi-level graphical modeling, to our best knowledge, has not been addressed by existing work yet. 

We propose a novel statistical method, the random covariance model, for bi-level graphical modeling which simultaneously learns the group- and individual-level graphical models. The proposed method assumes that each sub-dataset has a unique graphical model, which is a variant of a common unknown group-level graphical model featuring the shared correlation pattern. We utilize a penalty on the Kullback-Leibler (KL) divergence between the corresponding individual precision matrices and the group-level one, bringing similarity in estimating individual graphical models and pooling information contained in sub-datasets for group-level graphical model inference at the same time. Combined with $L_1$ penalties to foster sparsity in the estimators, the random covariance model leads to sparse precision matrix estimation at both the group and individual levels. Our proposed random covariance model accounts for the similarity between multiple individual-level graphical models with common correlation features, identifies the group-level network structures underlying the individual graphs in the tested group, and at the same time infers multiple individual-level networks by allowing for differences among them. 

Compared to existing multiple graphical modeling methods that focus on the individual level only, our proposed method has the following major contributions: (1) It simultaneously estimates the multiple individual-level as well as the group-level graphs, thus identifying unique structures in the individual-level connections while also being able to capture the shared correlation pattern at the group level; (2) It utilizes a penalty term on the KL divergence between the individual- and group-level covariance matrices to encourage similarity among individual-level models and estimate the group-level model, which can be nicely interpreted as a random covariance model that mimics the random effects model for the mean structure in linear regression; (3) The KL loss-based penalized likelihood method is computationally efficient to be able to scale up to datasets with a large number of individual-level networks to be inferred, as often seen in multi-subject neuroimaging or genetics data; (4) The interpretation as a random covariance model allows us to evaluate the complexity of the random covariance model by estimating the degrees of freedom in a way that is similar to the approach of Hodges and Sargent (2001) for a random effects model, which takes into account the bi-level covariance structure. This is practically useful for selecting the tuning parameters of the random covariance model. 

The outline for the rest of the paper is as follows. In Section \ref{sec:model}, we present our random covariance model for bi-level graphical modeling, a computational algorithm, and selection of tuning parameters. We present the asymptotic properties of our proposed methods in Section \ref{sec:theory}. We report results from our simulation study in Section \ref{sec:simu}, and apply the method to a real resting-state functional magnetic resonance imaging (fMRI) dataset for connectivity network inference in Section \ref{sec:case}. We finally conclude with a discussion in Section \ref{sec:diss}.

\section{Methodology}  \label{sec:model}

\subsection{The random covariance model} 

Suppose we have $K$ sub-datasets, $\bY^{(1)},\ldots, \bY^{(K)}$, in which each $\bY^{(k)}$ is an $n_k \times p$ matrix containing $n_k$ observations of a common $p-$dimensional random vector, $\by^{(k)}_i=(\by^{(k)}_{i1}, \ldots, \by^{(k)}_{ip})$, with $k=1,\ldots,K$ and $i=1,\ldots ,n_k$. We assume that the $\sum_k^K n_k$ observations are independent, and observations from each sub-dataset are identically distributed from a multivariate Gaussian distribution, i.e. $\by^{(k)}_i\sim N_p (\mu_k,\Sigma_k)$, where $\mu_k \in \mathtt{R}^p$ and $\Sigma_k$ is a positive definite $p\times p$ matrix. Without loss of generality, we assume the observations for each sub-dataset are centered such that $\mu_k=0$. 

We assume that there is a unique graphical model $G_k$ associated with each sub-dataset $k$, in which the nodes are the $p$ random variables, and two nodes $j$ and $j'$ are connected with an edge in $G_k$ if their corresponding element in the precision matrix $\Omega_k=\Sigma_k^{-1}$ is nonzero. Thus the problem of learning the graph $G_k$ is equivalent to estimating the covariance or precision matrix. We further assume that these $K$ graphical models are similar to each other and they are all variants from a common graphical model $G_0$, which can be considered as the group-level graphical model representing the shared connection pattern.

Let $S_k= \left( \bY^{(k)} \right)^T \bY^{(k)}/n_k$ be the sample covariance matrix of the $k\th$ sub-dataset, which is the maximum likelihood estimator (MLE) of $\Sigma_k$. When $p \gg n_k$, the sample covariance is often singular and thus cannot be inverted to yield an estimate of $\Omega_k$. A general approach to obtain more stable estimators of $\Omega_1,\ldots,\Omega_K$ is to minimize the objective function taking the form
\be
\sum_{k=1}^K \; \left\{ -\log\det(\Omega_k) + \mathtt{tr}(S_k\Omega_k) \right\}  + P(\Theta) \label{eq:ploglik}
\ee
subject to the positive definite constraint on $\Omega_1,\ldots,\Omega_K$. This is a penalized log-likelihood function composed of a negative log-likelihood function plus a penalty term $P(\Theta)$, with $\Theta$ denoting the set of parameters. In particular, the GLasso method specifies an $L_1$ penalty on $\Omega_k$, and yields a sparse precision matrix estimate and implied graphical model. In the context of multiple graphical modeling, Guo et al. (2011) proposes a hierarchical penalty that targets the removal of common zeros in the $\Omega_k$, and the JGL method applies $L_1$ penalties to $\Omega_k$'s and their pairwise differences, which achieves estimators of all $G_k$ with similar structures. 

In this paper, we consider inference of both the individual $G_k$ and the underlying group-level graphical model $G_0$. In particular, we propose a method with the penalty function $P(\Theta)$ in (\ref{eq:ploglik}) taking the form 
\be
P(\{ \Omega_k \}, \Omega_0) = \underbrace{\lambda_1 \sum_{k=1}^K |\Omega_k|_1 }_{P1}
			+ \underbrace{\lambda_2 \sum_{k=1}^K \left\{ -\log\det(\Omega_k\Omega_0^{-1}) + \mathtt{tr}				  (\Omega_k \Omega_0^{-1}) - p)\right\} }_{P2}
			+ \underbrace{\lambda_3 |\Omega_0|_1 }_{P3}    \label{eq:penalty}
\ee
where $\lambda_1$, $\lambda_2$, and $\lambda_3$ are non-negative tuning parameters. Here we introduce a positive definite matrix $\Omega_0$ in the penalty function, of which the nonzero off-diagonal structure is assumed to give the group-level graphical model $G_0$.  We can consider $\Omega_0$ as the group-level precision matrix corresponding to the overall distribution of pooled data. Thus the penalty function is composed of three parts: $(P1)$, $L_1$ penalties on $\Omega_k$ to induce sparsity in the individual graphical models $G_k$; $(P2)$, a penalty on the KL-divergence between each individual $\Omega_k$ and $\Omega_0$ associated with the underlying group-level graph $G_0$; and $(P3)$, an $L_1$ penalty on $\Omega_0$ to induce sparsity in the group-level graphical model $G_0$. 

The KL-divergence can be considered as a measure of the distance between the covariance/precision matrices of two Gaussian distributions. By penalizing on the KL-divergence between each $\Omega_k$ and $\Omega_0$ as in $(P2)$, the proposed method actually shrinks all individual precision matrices to the group-level precision matrix $\Omega_0$. By combining the three penalty terms, we (1) obtain a group-level graph $G_0$ by pooling information from all individual graphs $G_k$, and (2) estimate each $G_k$ by borrowing strength from other sub-datasets through the group-level $\Omega_0$. Compared to existing multiple graphical models such as Guo's method and the JGL, our proposed method estimates the underlying group-level structure as well as the multiple individual graphs, and is computationally efficient to scale up to large datasets that involve many subjects or conditions. The efficiency roots in the fact that the computation is linear in $K$ and can be parallelized due to the independence of the $K$ individuals given the group-level $\Omega_0$. Note that from a Bayesian point of view, this penalty term $(P2)$ can be represented as independent Wishart distributions of $\Omega_k$ that are centered at $\Omega_0$ with a degrees of freedom (df) $\lambda_2$. Thus the proposed method can be interpreted as a random covariance model, in which each subject-level covariance is a random level from an inverse Wishart distribution centered at the group mean.

\subsection{Computational Algorithm}

The objective function (\ref{eq:ploglik}) with penalty (\ref{eq:penalty}) is not convex. We use a block coordinate descent (BCD) algorithm for maximizing it, which iteratively updates the two blocks, $\{ \Omega_k \}_{k=1,…,K}$ and $\Omega_0$, respectively. Specifically, the BCD algorithm works as follows:
\begin{enumerate}

\item Initialize $\widehat\Omega_k=(1-\rho)S_k+\rho I_k$ for $k=1,\ldots, K$, and $\widehat\Omega_0= \sum_k \Omega_k /K$, where $\rho$ is a small value.
\item For $k=1,\ldots,K$, update $\widehat\Omega_k$ by solving 
\be
{\arg\min}_{\Omega_k } \left\{ -\log\det \Omega_k + \mathtt{tr}\left( \frac{S_k+\lambda_2 \widehat\Omega_0^{-1}}{1+\lambda_2} \Omega_k \right) + \frac{\lambda_1}{1+\lambda_2} |\Omega_k |_1 \right\} \label{eq:step2}
\ee
\item Update $\widehat\Omega_0$ by solving 
\be
{\arg\min}_{\Omega_0 } \left\{ \log\det\Omega_0 + \mathtt{tr}\left( \frac{ \sum_{k=1}^K \widehat\Omega_k}{K} \Omega_0^{-1} \right) + \frac{\lambda_3}{K\lambda_2} |\Omega_0 |_1 \right\}  \label{eq:step3}
\ee
\item Repeat Steps 2 and 3 until convergence is achieved.

\end{enumerate}
	
Note that in Step 2, each $\Omega_k$ can be solved independently using the graphical lasso method, in which the typical sample covariance matrix is replaced by a weighted average of the $k\th$ sample covariance and the current estimator of the group-level covariance matrix $\widehat{\Omega}_0^{-1}$. Step 3 parallels the algorithm of estimating a sparse covariance matrix (Bien and Tibshirani, 2011).  The objective function is non-convex, which decomposes into the sum of a convex and concave function. Bien and Tibshirani (2011) utilizes a majorize-minimize iteration to solve (\ref{eq:step3}), while Wang (2012) developed the coordinate descent algorithm and the Expectation/Conditional maximization algorithm for minimizing (\ref{eq:step3}). We follow Wang (2012) to use the coordinate descent algorithm to solve (\ref{eq:step3}), which, in joint with Step 2, leads to a BCD algorithm. While we cannot guarantee to yield a global minimizer of the non-convex problem, the limiting points of such an algorithm will be local minimizers that are critical points of the objective function (An and Tao, 2005).

It is  noted that since each $\Omega_k$ can be solved independently in Step 2 given a current estimate of $\Omega_0$, estimation of $\Omega_k$  can be conducted in parallel at each iteration and the total computing time is only linear in the number of sub-datasets $K$. This makes our method computationally scalable to high-dimensional data with a large value of $K$.

\subsection{Tuning parameter selection}  \label{subsec:tune_bic}

Commonly used methods including penalized likelihood approaches and cross validation can be applied to select the tuning parameters $\blambda=(\lambda_1, \lambda_2, \lambda_3)$. In the context of high dimensionality as we see in the problem of multiple graphical model inference, Bayesian information criterion (BIC) is a commonly accepted choice for tuning parameter selection. The BIC formula for the random covariance model is given by
\bse
\text{BIC}^1(\blambda) = \sum^K_{k=1} \left[ tr\left( S_k \widehat{\Omega}_k(\blambda) \right) - \log \det\left( \widehat{\Omega}_k(\blambda) \right) + df_k \log(n_k) 	\right] ,
\ese
where $\widehat{\Omega_k}(\blambda)$ is the estimated precision matrix for the $k\th$ sub-dataset with the tuning parameters $\blambda=(\lambda_1,\lambda_2,\lambda_3)$, and the degrees of freedom $df_k$ are defined as the number of nonzero off-diagonal elements in $\widehat{\Omega}_k$. 

However, the above BIC criterion considers the $K$ sub-datasets separately and ignores the hierarchical structure in our concerned problem and the similarity across the $K$ individual covariances. Here we use a BIC criterion that is based on a definition of the degrees of freedom for the random covariance model, which accounts for the hierarchical structure of the common group-level covariance and random individual covariances. In particular, we define the degrees of freedom when $n_1=\ldots=n_K=n$ as 
\be
df = \left(\sum_k \frac{df_k}{1+\lambda_2}+\frac{\lambda_2 df_0}{1+\lambda_2} \right), \label{eq:dof}
\ee
where $df_k$ and $df_0$ are the number of nonzero off-diagonal elements in $\widehat{\Omega}_k$ and $\widehat{\Omega}_0$, respectively.

The above formula for degrees of freedom is similar to that proposed by Hodges and Sargent
(2001) for the degrees of freedom of a random effects model. It relies on the tuning parameter $\lambda_2$, which controls the strength of penalty on the KL divergence between $\Omega_k$ and $\Omega_0$. From the Bayesian perspective, $\{ \Omega_k \}$ can be considered as random levels of $\Omega_0$ in the random covariance model, and $\lambda_2$ controls the extent of shrinking $\Omega_k$ toward $\Omega_0$ and thus the degrees of freedom allocated to each $\Omega_k$. In fact, $\lambda_2/(1+\lambda_2)$ is proportional to the amount of uncertainty controlled by the group-level covariance/precision matrix, and $1/(1+\lambda_2)$ approximates the proportion of uncertainty controlled by each individual precision matrix. When $\lambda_2 \rightarrow 0$, no shrinkage is imposed and each level can be considered as independent with $df=\sum_k df_k$; when $\lambda_2 \rightarrow \infty$, we have $\Omega_k=\Omega_0$ for all $k$ with $df=df_0$.

Based on the degrees of freedom defined in (\ref{eq:dof}), we can use the following BIC criterion for selecting the tuning parameters for the random covariance model:
\bse
\text{BIC}^2(\blambda) &=& \sum^K_{k=1} \left[ tr\left( S_k \widehat{\Omega}_k(\blambda) \right) - \log \det\left( \widehat{\Omega}_k(\blambda) \right) 	\right] 
 +  df \log\left( Kn \right) .
\ese 
Our simulation studies show that this BIC works well to select group- and individual-level precision matrices with high true positive rates and low false positive rates for our proposed random covariance model.

\section{Asymptotic properties} \label{sec:theory}

In this section, we present some asymptotic properties of the proposed random covariance model when $n_1=\ldots=n_K=n$. Let $\Omega_k$ be the precision matrix of the $k\th$ sub-dataset and $E_k=\{(j,j^\prime): j \leq j^\prime, \omega^{k}_{j,j^\prime} \neq 0 \}$ be the set of indices of nonzero off-diagonal elements in $\Omega_k$. Let $\Omega_0=\sum^K_{k=1}\Omega_k/K$, and $q=|E_0|=|E_1 \cup \cdots E_K|$ be the cardinalities of $E_0$ which is the union of $E_k$. We assume that the following regularity conditions hold:
\begin{itemize} \setlength\itemsep{-1em}
\item[A1.] There exist constants $\tau_1$ and $\tau_2$ such that 
$$ 0 < \tau_1 < \psi_{\min}(\Omega_k) \leq \psi_{\max}(\Omega_k) < \tau_2 < \infty \quad \text{ for all } k=1,\ldots,K $$

\item[A2.] There exists a small positive constant $\epsilon > 0$  such that 
$$||\Omega_k - \Omega_0|| < \epsilon \quad \text{ for all } k=1,\ldots,K $$
\end{itemize}

Condition A1 bounds uniformly the eigenvalues of $\Omega_k$, which is standard for covariance/precision matrices as in Bickel and Levina (2008) and Lam and Fan (2009). It guarantees that the precision matrices exist and are well-conditioned. Condition A2 bounds the operator norm of the differences between individual precision matrices and their group mean.

\vspace{.2in}
Theorem 1:  {\it 
Suppose Conditions A1 and A2 hold, if $\lambda_1 \asymp \{\log p/n\}^{1/2}$, $\lambda_2=O(\{\log p/n\}^{1/2})$, and $\lambda_3/\lambda_2=O(\{\log p/n\}^{1/2})$, then there exists a local minimizer $(\widehat{\Omega}_0, \{ \widehat{\Omega}_k \}_{k=1}^K)$ such that $ \sum_{k=1}^K ||\widehat{\Omega}_k-\Omega_k||_F= O_P(\{(q+p)logp/n\}^{1/2}) $ and $ ||\widehat{\Omega}_0-\Omega_0||_F= O_P(\{(q+p)logp/n\}^{1/2}) $.
}

\vspace{.2in}
Theorem 2: {\it 
Suppose all the conditions in Theorem 1 hold, and further assume that the local minimizer $(\widehat{\Omega}_0,\{\widehat{\Omega}_k\}_{k=1}^K)$ in Theorem 1 satisfies $\sum_{k=1}^K ||\widehat{\Omega}_k-\Omega_k||^2=O_P(\eta_n)$ and $||\widehat{\Omega}_0-\Omega_0||^2=O_P(\eta_n)$ for a sequence of $\eta_n \rightarrow 0$. If $\{\log p/n\}^{1/2}+\eta_n^{1/2}=O(\lambda_1)$ and $\eta_n^{1/2}=O(\lambda_3/\lambda_2)$, then with probability tending to 1, $\hat{\omega}^k_{jj^\prime}=0$ for all $(jj^\prime) \in E^c_k$ and $k=1,\ldots, K$, and $\hat{\omega}^0_{jj^\prime}=0$ for all $(jj^\prime) \in E^c_0$.
}

\vspace{.2in}
The proofs are provided in the Appendix. Theorem 1 ensures the consistency of both the individual precision estimates and the group-level estimates. Compared to independent graphical lasso, it requires extra upper bounds on $\lambda_2$ and $\lambda_3/\lambda_2$. Theorem 2 indicates that the sparsistency requires a lower bound on $\lambda_1$, which is similar to independent graphical lasso, for the individual precision estimators, as well as an extra lower bound on $\lambda_3/\lambda_2$ for the group-level estimator.

\section{Simulation study}  \label{sec:simu}

In this section, we examine the finite-sample performance of our proposed random covariance model using simulations. We evaluate the performance in individual network inference, group-level network inference, as well as the computational times when $n << p$. We also include three competitive methods in the simulations for comparison, the independent graphical lasso, Guo's multiple graphical models, and the joint graphical lasso (JGL). Note that there are no existing methods for bi-level graphical modeling to our knowledge. All these three competitive methods only obtain estimates of the individual-level graphical models. In addition, considering the heavy computational burden of the latter two methods, we restricted our simulation setting to a relatively small value of $K$. However, our method can scale up to datasets with a much larger $K$, which is often seen in real applications of multi-subject data, for example the fMRI data in our case study.

\begin{figure}[h]
\centerline{\includegraphics[width=6in]{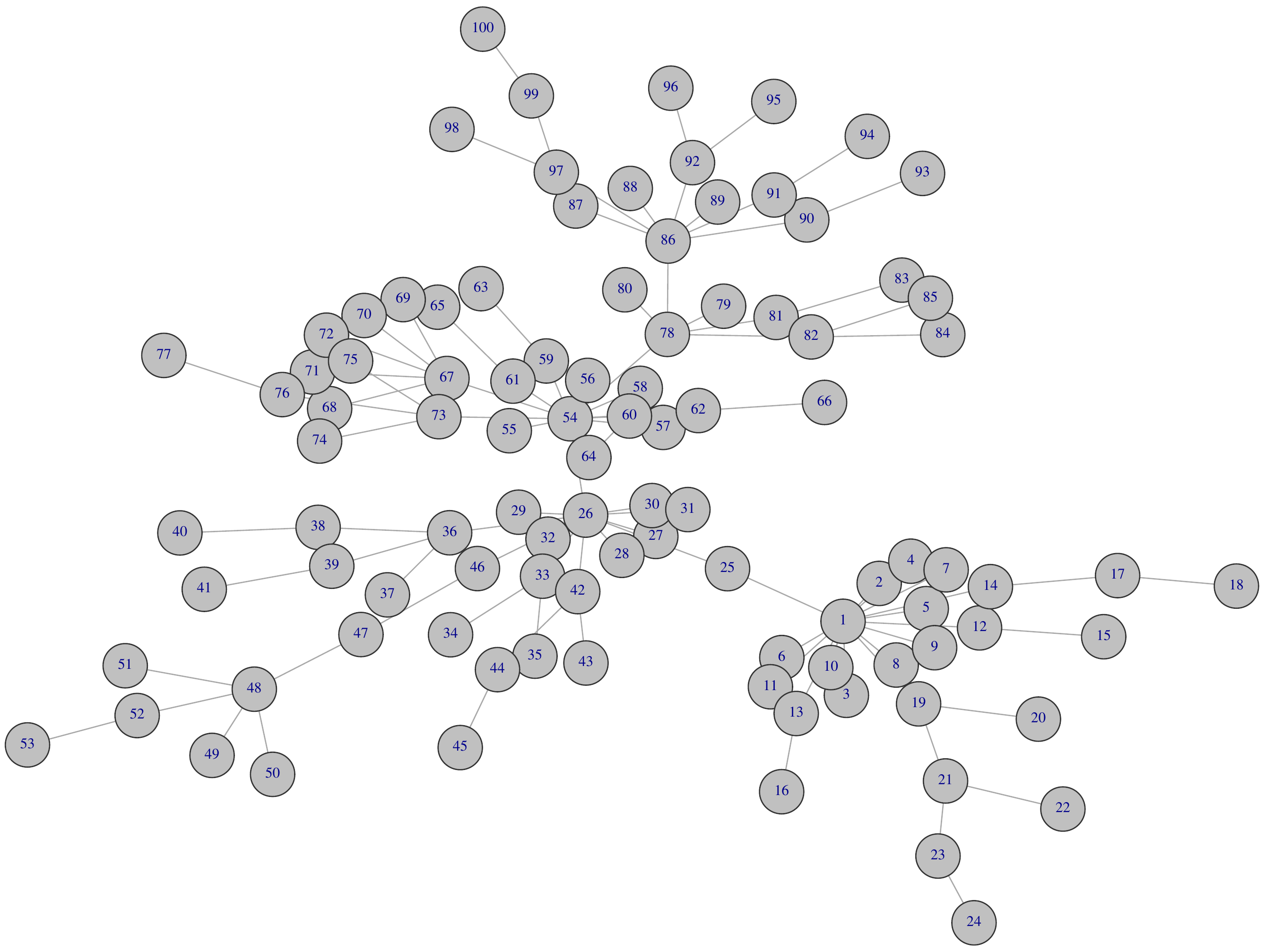}}
\caption{The group-level network in the simulation studies.}
\label{fig:simu_network}
\end{figure}

We considered three scenarios to generate the data, all assuming a common  group-level network structure with $p=100$ nodes as shown in Figure \ref{fig:simu_network}. Given the network structure, we generated the group-level precision matrix as follows. We first created a $p \times p$ identity matrix. Then for elements corresponding to edges in the network, we generated random values from a uniform distribution with support on $\{ [-1,-0.5] \cup [0.5,1] \}$. To ensure positive definiteness, we finally divided each off-diagonal element by the total number of non-zero elements of that row.

We then generated $K=8$ individual network structures by randomly picking $\rho M$ pairs of nodes in the graph and adding/removing edges to/from the group-level graph, where $M$ is the number of edges in the group-level network, and $\rho$ is the ratio of number of differential edges in individual networks to the number of edges in the group-level network. In the simulations, we considered three scenarios with $\rho=0, 0.2,$ and $0.4$ respectively, reflecting different levels of group-vs-individual similarity. The individual precision matrices were generated the same as the group-level precision matrix, but with the values of the corresponding elements of the added edges generated from the intervals $[-1,-0.5] \cup [0.5,1]$ and those of the deleted edges to be set as zero. Finally, we generated a sub-dataset of sample size $n_k=50$ from a zero-mean Gaussian distribution given each individual precision matrix. We generated 100 datasets for each scenario as described above, and applied the three competitive methods to each dataset with varying tuning parameters. 

\begin{figure}[h]
\centering
\vspace{1in}
\rotatebox{90}{
\begin{minipage}{\columnwidth}
\centerline{\includegraphics[width=9in,height=5in]{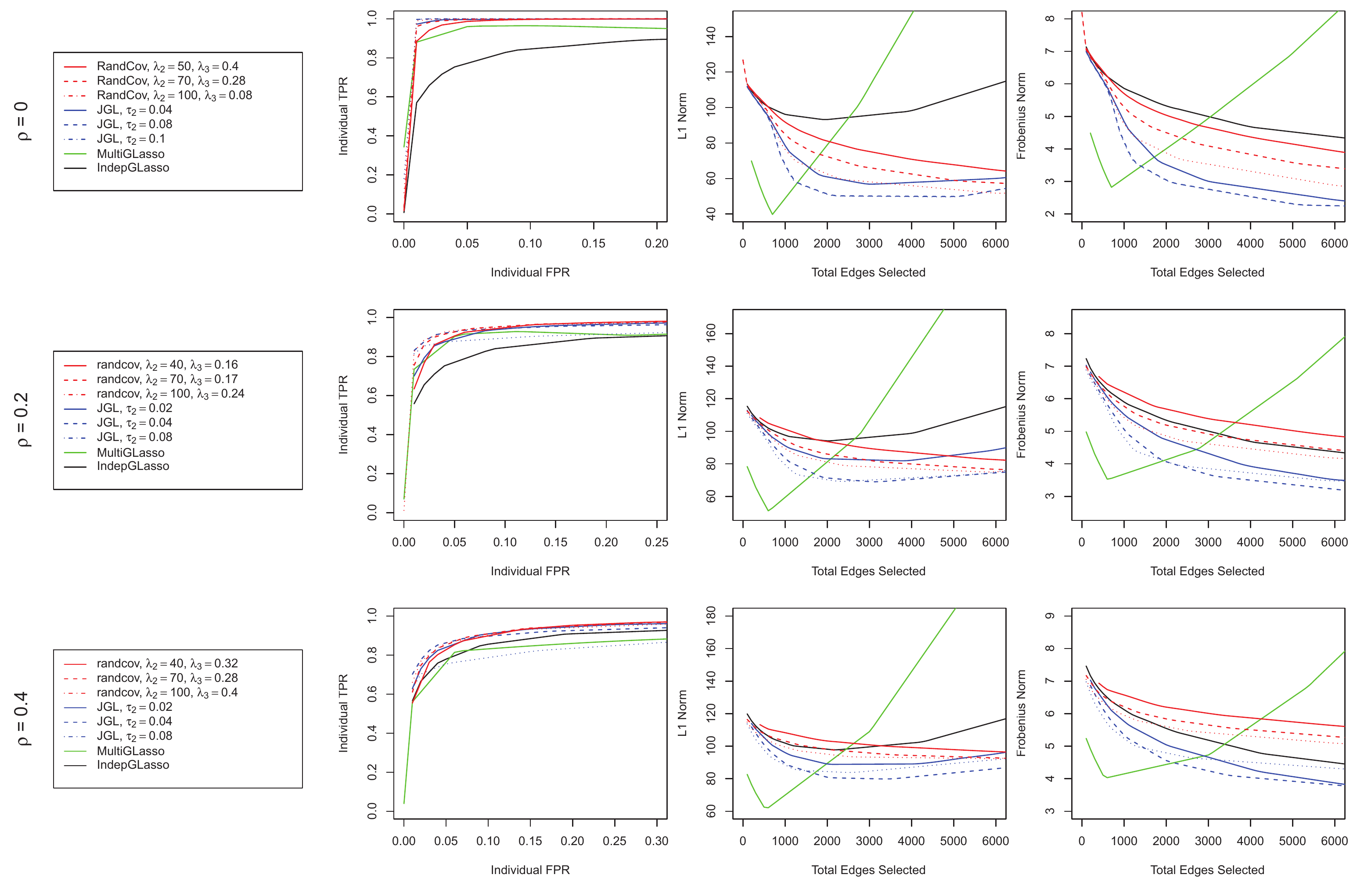}}
\caption{The performance of the methods in individual-level graphical modeling for the scenarios of $\rho=0$ (upper panel), 0.2 (middle panel), and 0.4 (lower panel) respectively. The red curves correspond to the random covariance model, the blue curves the JGL method, the green curves Guo et al's multiple graphical model, and the black curves the independent graphical lasso method. Each curve is obtained by averaging across 100 replicates. } 
\label{fig:sim1}
\end{minipage}}
\end{figure}

Figure \ref{fig:sim1} shows the performance of the four methods in individual-level graphical modeling averaged across the 100 replicates for each of the scenarios of $\rho=0, 0.2,$ and $0.4$ respectively. The red curves correspond to our proposed random covariance model, for which we fixed $\lambda_2$ and $\lambda_3$ at different values and displayed the curves of measures with varying values of $\lambda_1$. The blue curves correspond to the JGL method, for which $\lambda_2$, the parameter for penalizing pairwise differences, was fixed and $\lambda_1$, the parameter for sparsity, was varied. The green and black curves correspond to Guo et al's multiple graphical models and the independent graphical lasso method, respectively, with their tuning parameter varied. 

Figure \ref{fig:sim1} the left panel displays the true positive rates (TPRs) versus the false positive rates (FPRs) regrading edge identification for the individual-level graphical models. The results indicate that the random covariance model has similar performance to the JGL method and better performance than Guo et al's and the independent graphical lasso methods, especially when the individual-level graphical models have a high degree of similarity. The middle panel displays the mean $L_1$ norms of the differences between individual precision estimates and their true values as the inferred number of edges increases. Guo et al's method has the lowest error when the estimated matrices are extremely sparse, but the errors surge rapidly when the precision estimates become denser. For the other three methods, the JGL has the best performance and the random covariance model is between the JGL and the independent graphical lasso. The right panel of Figure \ref{fig:sim1} displays the mean frobenius norms of the estimation errors for individual-level precision matrices as the inferred number of edges increases. The results are similar to those of the $L_1$ norms but the random covariance model has greater loss to the JGL and less obvious gain to the independent graphical lasso. These indicate that our proposed random covariance model works well as a model selection for individual-level graphical model inference but introduces higher biases in estimating individual precision matrices by shrinking them toward the group-level precision estimator.

\begin{figure}[h]
\vspace{1in}
\centering
\rotatebox{90}{
\begin{minipage}{\columnwidth}
\centerline{\includegraphics[width=7in,height=5in]{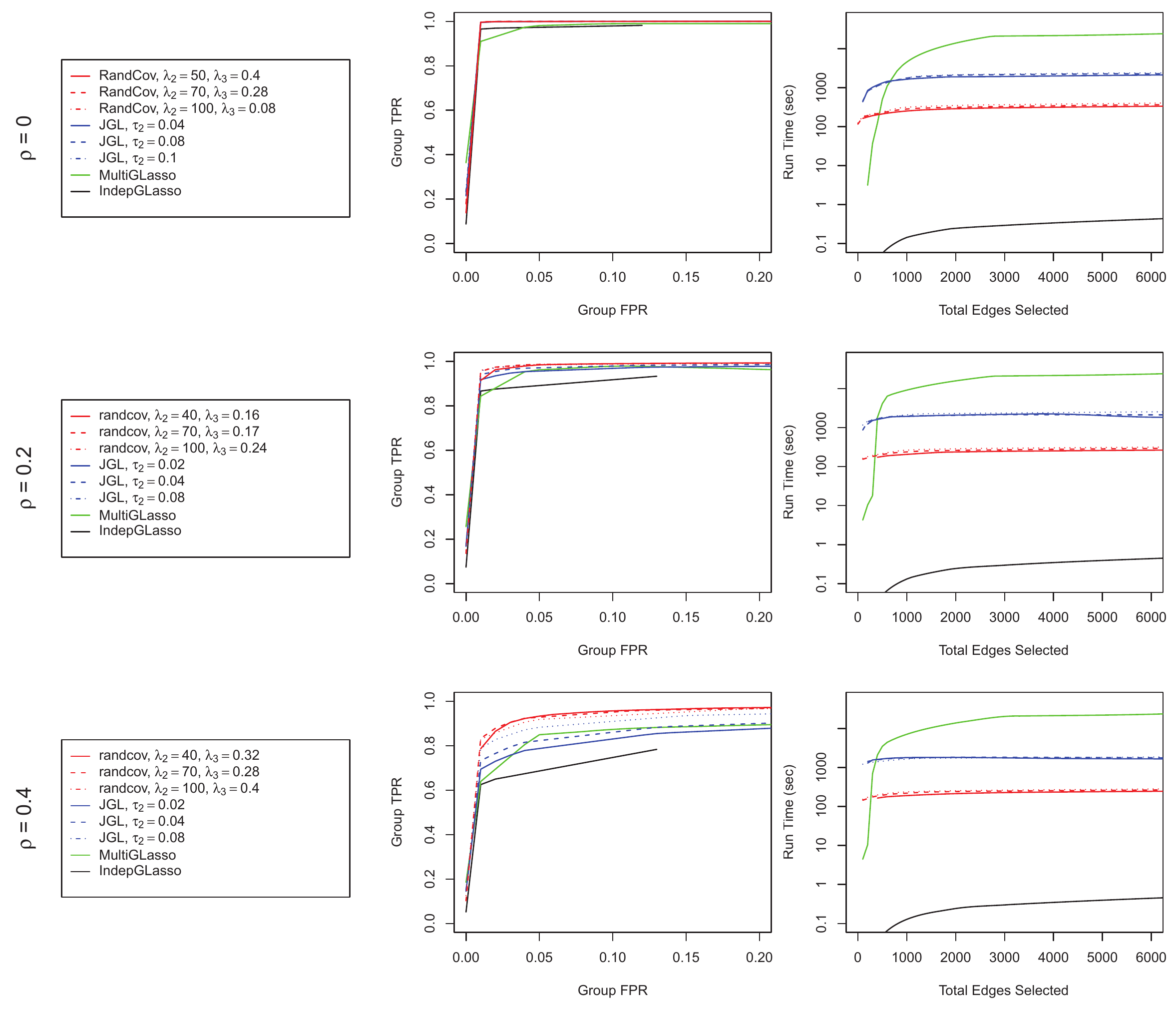}}
\caption{The performance of the methods in group-level edge identification and computational time for the scenarios of $\rho=0$ (upper panel), 0.2 (middle panel), and 0.4 (lower panel) respectively. The red curves correspond to the random covariance model, the blue curves the JGL method, the green curves Guo et al's multiple graphical model, and the black curves the independent graphical lasso method. Each curve is obtained by averaging across 100 replicates. }
\label{fig:sim2}
\end{minipage}}
\end{figure}

Figure \ref{fig:sim2} the left panel displays the average TPRs versus the FPRs in terms of edge identification for the group-level graphical model obtained by the proposed random covariance model. For comparison, we also include in the figure the curves obtained from the other three methods. As these competitive methods do not make inference at the group level, we somewhat arbitrarily define their estimated group-level network such that an edge is included in the group-level network if the edge is present in more than half of the estimated individual-level networks. We observe that our random covariance model has the best performance in the group-level edge identification for all the scenarios, and the gain is more obvious when the individual graphs deviate further from the common group-level graphical model. 

The right panel of Figure \ref{fig:sim2} presents the computational time (in seconds) of the four methods. We see that with $K=8$, the run time of the random covariance model is about 1/10 that of the JGL method and about 1/100 that of Guo et al's method. We can easily extrapolate that the gain of our method in computational efficiency will be even bigger with larger number of individual levels.

For practical use, we need to choose one model with some criterion. Table \ref{tab:sim_bic} presents the performance of the inferred models by the four methods with the tuning parameters selected by the BIC values. We present two random covariance models in the table. One was selected by BIC$^1$ and the other by BIC$^2$ as described in Section \ref{subsec:tune_bic}. The models of the JGL, Guo et al's multiple graphical model, and independent graphical lasso methods, were all selected by BIC$^1$. We can observe that the combination of the random covariance model and the BIC$^2$ criterion we introduced in Section \ref{subsec:tune_bic} results in inferred bi-level graphical models with best balanced performance in group- and individual-level graphical model estimation for all the three scenarios with different levels of group-vs-individual similarity.

\begin{table}[h]
\centering
\begin{tabular}{ccccccccc}
\hline
\multicolumn{2}{c}{} & {BIC} & {ITPR} & {IFPR} & {GTPR} & {GFPR} & {Frobenius} & {$L_1$ norm} \\
\hline
\multirow{5}{*}{\rotatebox{90}{$\rho=0$}} & {RandCov1} & 47963$^*$ &  0.9994 & 0.0394 & 0.9993 & 0.0128 & 3.6788 & 59.6208 \\
& {RandCov2} & 55294  &  0.9731 & 0.0039 & 0.9721 & 0.0017 & 4.9107 & 78.6814 \\
& {JGL}  	& 54022 & 0.9991 & 0.0112 & 0.9992 & 0.0106 & 3.5993 & 58.1425 \\
& {Multi}  	& 49296 & 0.9613 & 0.0461 & 0.9880 & 0.0385 & 4.5624 & 95.6536 \\
& {Indep}  	& 57217 & 0.5138 & 0.0063 & 0.4442 & 0.0000 & 6.2553 & 100.7836 \\
\hline
\multirow{5}{*}{\rotatebox{90}{$\rho=0.2$}} & {RandCov1} & 50410 & 0.9617 & 0.1167 & 0.9841 & 0.0361 & 4.2911  & 75.1677\\
& {RandCov2} & 57471 & 0.7808 & 0.0065 & 0.8128 & 0.0009 & 5.5880 & 91.8405 \\
& {JGL}  	& 55242 & 0.8401 & 0.0104 & 0.8492 & 0.0014 & 4.9358 & 81.6280 \\
& {Multi}  	& 50157 & 0.7103 & 0.0018 & 0.8147 & 0.0002 & 3.4391 & 49.4998 \\
& {Indep}  	& 57516 & 0.5279 & 0.0078 & 0.3113 & 0.0000 & 6.2346 & 101.4076\\
\hline
\multirow{5}{*}{\rotatebox{90}{$\rho=0.4$}} & {RandCov1} & 52995$^*$ & 0.9201 & 0.1125 & 0.8719 & 0.0194 & 5.2777 & 92.3807 \\
& {RandCov2} & 59250 &  0.3535 & 0.0019 & 0.2725 & 0.0000 & 6.5436 & 107.7751 \\
& {JGL}  	& 56310 & 0.7378  & 0.0120 & 0.5723 & 0.0008 & 5.2698 & 88.4216 \\
& {Multi}  	& 51459 & 0.5572  & 0.0032 & 0.6029 & 0.0005 & 3.9745 & 59.9311 \\
& {Indep}  	& 58558 & 0.5567  & 0.0093 & 0.1719 & 0.0000 & 6.3779 & 104.6595 \\
\hline
\end{tabular}
\caption{Performance of the models selected by BICs averaged over 100 replicates. RandCov gives the random covariance model with the tuning parameters selected by BIC$^2$;
RandCov2 and other methods give the model selected by BIC$^1$. ITPR: individual-level true positive rate; IFPR: individual-level false positive rate; GTPR: group-level true positive rate; GFPR: group-level false positive rate; Frobenius: mean frobenius norm of individual precision estimation errors; $L_1$ norm: mean $L_1$ norm of individual precision estimation errors. Symbol $*$ indicates BIC values defined by BIC$^2$ while other BIC values are by BIC$^1$.} \label{tab:sim_bic}
\end{table}

\section{Functional connectivity analysis of fMRI data}  \label{sec:case}
 
Schizophrenia is a serious mental disorder characterized by a lack of integration between thought, emotion, and behavior. The pattern of functional connectivity in schizophrenia is of interest to help determine whether functional connectivity disruptions play a role in the lack of integration of information processing. We applied the random covariance model to a resting-state fMRI dataset collected from schizophrenia patients. The fMRI data were collected from 16 first-episode schizophrenia patients. Each patient underwent a 6-min resting-state fMRI scan with a total of 180 volumes of images collected, each containing measurements at $64 \times 64 \times 34$ voxels. See Camchong et al. (2011) for detailed characteristics of imaging data. For our data analysis, we only focused on the 60 volumes in the last one third of the session which appeared to be more stable in our data exploration. 

We reduced the dimension by parcellating the brain into 120 ROIs using the Automated Anatomical Labeling (AAL) (Tzourio-Mazoyer et al., 2002) and extracting the mean measures for each ROI. Four regions were removed due to abnormally high variances, which left us with a total of 116 ROIs. The ROI-level data were then processed such that the resulting 16 sub-datasets were all centered at zero and had the same total variance. We applied the random covariance model to the data, aiming to infer the underlying network of functional connectivity among the ROIs shared by the group of patients as well as the unique network structures of each patient in the data at the same time. The modified BIC$^2$ was used to choose the tuning parameters. For comparison, we also applied the graphical lasso method to each subject's fMRI data, which infers the functional connectivity network of each patient separately. The JGL and Guo's methods were also tried for individual-level network inference, but ran out of time for the fMRI data due to the high dimensionality and large number of subjects. 

\begin{figure}[h]
\centerline{\includegraphics[width=5in,height=5in]{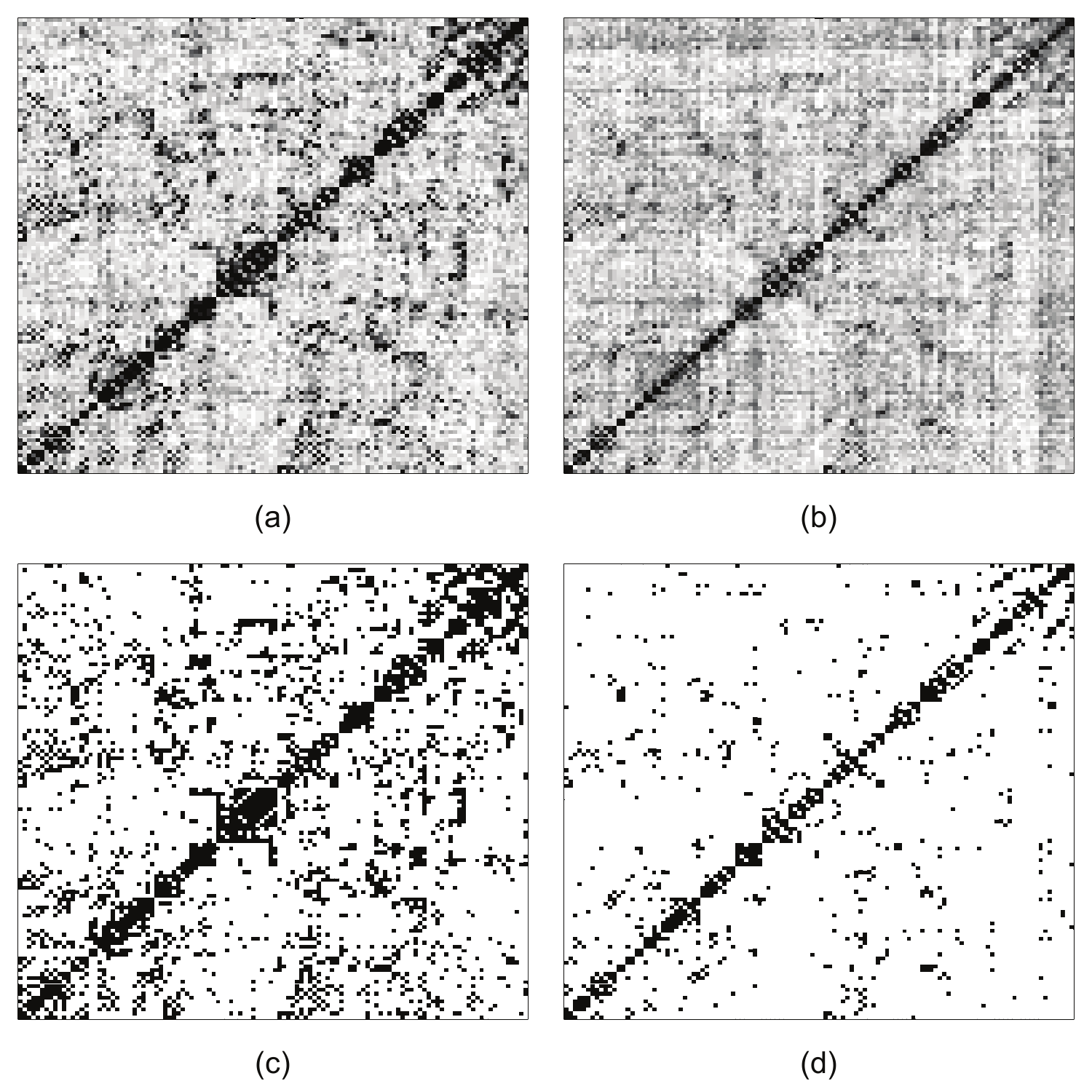}}
\caption{Image plots of inferred networks by the random covariance model and independent graphical lasso. (a) The mean adjacency matrix inferred by the random covariance model averaged across the 16 subjects; (b) The across-subject mean adjacency matrix inferred by the independent graphical lasso; (c) The group-level network inferred by the random covariance model; (d) The common network by the independent graphical lasso.}
\label{fig:real_image}
\end{figure}

Figure \ref{fig:real_image}(a \& b) summarizes the subject-specific graphical models inferred by the random covariance model and independent graphical lasso, respectively, using the mean adjacency matrices, which were obtained by averaging the individual graphical models across the 16 subjects. The darkness of each dot gives the proportion of the subjects who exhibit functional connectivity between the corresponding ROIs. The two images display similar patterns for the subject-level connection networks. However, the mean adjacency matrix of the random covariance model has higher contrast than that of the independent graphical lasso in the sense that the dark dots are more black and the light dots are more white. This indicates that the random covariance model borrows information across subjects, and consequently the inferred individual graphical models for each subject are more consistent with those for other subjects. 

\begin{figure}[h]
\centerline{\includegraphics[width=6in]{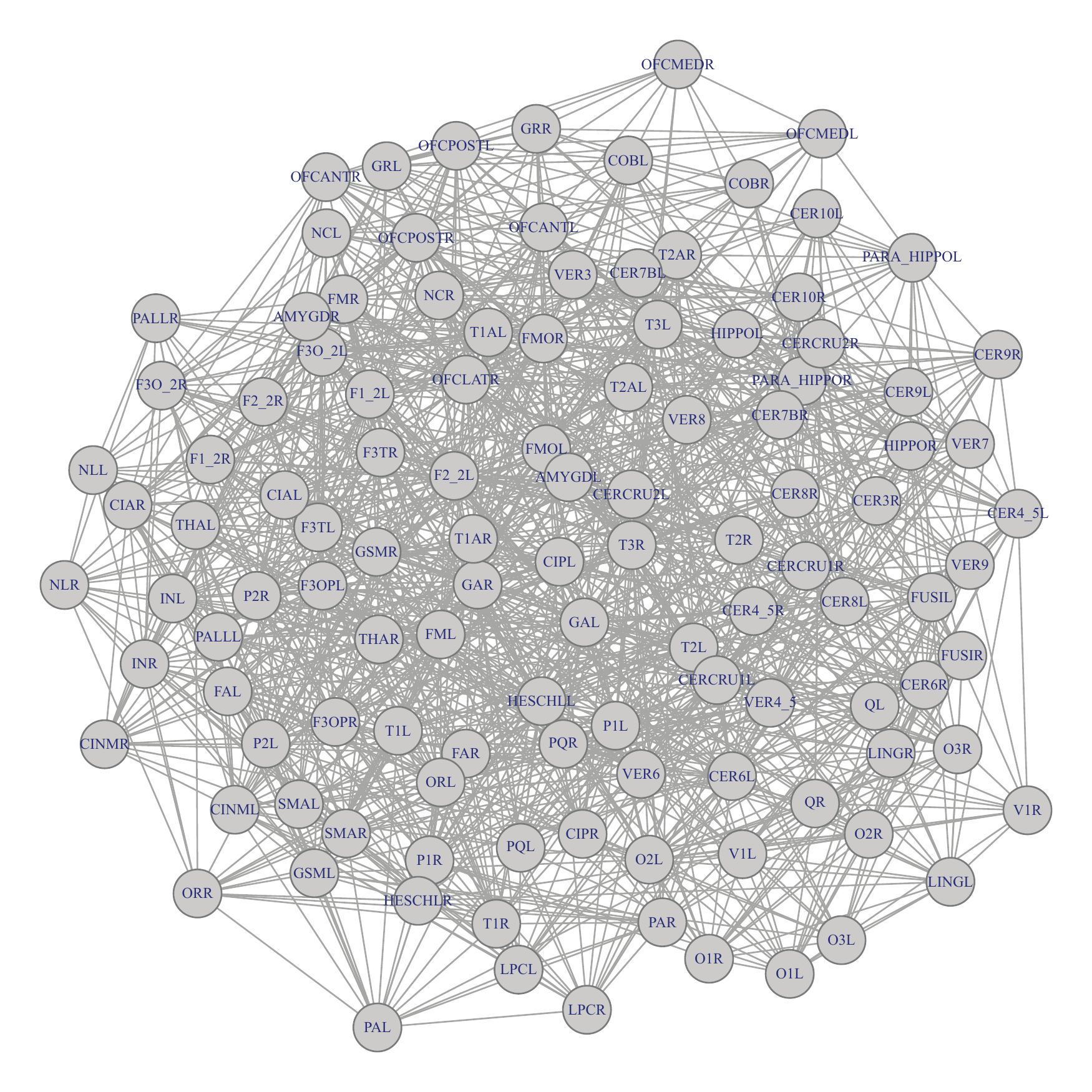}}
\caption{The group-level network of the 116 ROIs inferred by the random covariance model.}
\label{fig:real_network}
\end{figure}

Figure \ref{fig:real_image}(c) depicts the adjacency matrix for the inferred group-level graph with the corresponding network plot of the 116 ROIs displayed in Figure \ref{fig:real_network}. The observed pattern is consistent with the subject-level mean adjacency matrix in \ref{fig:real_image}(a), suggesting that the inferred group-level network includes the edges that are detected in most of the subject-specific networks. Thus the group-level graphical model can be considered to elicit the characteristics in functional connectivity that are shared by the group of schizophrenia patients. As no group-level graphical model is formally obtained by the independent graphical lasso, we present a group-level network for the method in which we arbitrarily chose to include edges that are shared by a majority of the subjects in the subject-specific networks, which is much more sparse. 


\section{Discussion}  \label{sec:diss}

We have proposed a random covariance model for inference of bi-level graphical models, which learns both group- and individual-level graphical models simultaneously. The method uses a KL loss-based penalty to shrink individual-level precision matrices to the group-level one, obtains estimates of group-level network structures shared by the group by pooling information from individuals, and identifies associations that are unique to individuals at the same time. The method can be nicely interpreted as a random covariance model that mimics the popular random effects model in linear regression. We further propose a measure of degrees-of-freedom that accounts for the hierarchical structure in the random covariance model, based on which a BIC is developed for model selection for the random covariance model in practical applications. The method is also computationally efficient and tractable to handle data with a large number of individual-level sub-datasets, which is often observed in real data analysis involving multiple subjects or conditions. 

We have demonstrated the asymptotic properties of the method and showed its finite-sample performance through simulation studies and real data application. The simulation results show that our method is able to automatically learn the group-level graph which extracts the shared characteristics of individual graphical models. At the individual level, our method has similar performance to the JGL method in individual-level edge identification. The method also enjoys computational efficiency that allows it to be applicable to datasets with a large number of sub-datasets, a feature notably attractive and useful in applications to multi-subject/condition data as we demonstrated in our functional connectivity analysis of the schizophrenia fMRI data. However, our method might be slightly worse than the JGL in the Frobenius and $L_1$- norm of errors in individual-level precision matrix estimation. This is expected since the JGL penalizes pairwise differences while our method shrinks all individuals to the group-level precision matrix.  Note that the JGL does not, while our method does, give a group-level estimate that identifies the shared correlation patterns among the group of individuals.

More advantageously, our model utilizes a KL penalty between individual- and group-level precision matrices which, from a Bayesian standpoint, can be represented as independent Wishart priors on individual precision matrices that center at the group-level mean. Thus the random covariance model developed in a frequentist framework can be easily transformed into a Bayesian form, which is a flexible alternative that can incorporate various complex features of data not easily incorporated in the frequentist approach. An example that is likely to be observed in real life is the presence of outliers at the individual level with the corresponding network structures deviating from the group-level one. We leave this to our next study. 

\section*{Acknowledgements}

This research is supported in part by NIH grant 1R03MH115300 and Grand-in-Aid of Research, Artistry and Scholarship provided by University of Minnesota (to L. Z., A. D., and K. Q.). The content is solely the responsibility of the authors and does not necessarily represent the official views of the National Institutes of Health or the University of Minnesota.

\section*{Appendix} 
We first present two lemmas that were established in Bickel and Levina (2008) and Lam and Fan
(2009), which will be used in the proof of Theorems 1 \& 2.

\vspace{.2in}
Lemma 1. {\it
Let $A$ and $B$ be real matrices such that the product AB is defined. Then, defining $||A||^2_{\min} = \psi_{\min}(A^TA)$, we have 
$$
||A||_{\min}||B||_F \leq ||AB||_F \leq ||A|| ||B||_F.
$$
In particular, if $A=(a_{jj^\prime})$, then $|a_{jj^\prime}| \leq ||A||$ for all $(j,j^\prime)$.
}

\vspace{.2in}
Lemma 2. {\it
Let $Z_i$ be i.i.d $\mathcal{N}(\0,\Sigma)$ and $\psi_{\max}(\Sigma) \leq \tau_1^{-1} \leq \infty$. Then, if $\Sigma=(\sigma_{ab})$,
$$
P\left[ \left| \sum_{i=1}^n(Z_{ij}Z_{ik}-\sigma_{jk}) \right| \geq nv \right] \leq c_1 \exp (-c_2nv^2) \text{ for } |v| \leq \delta,
$$
where $c_1$, $c_2$ and $\delta$ depend on $\tau_1$ only.
}

\vspace{.2in}
{\it Proof of Theorem 1.} For simplicity, we will use the following notations for the section: $\Omega=(\Omega_0,\left\{ \Omega_k \right\}_{k=1}^K )$, $\Delta=(\Delta_0,\left\{ \Delta_k \right\}_{k=1}^K )$, and $\Omega+\Delta=(\Omega_0+\Delta_0,\left\{ \Omega_k+\Delta_k \right\}_{k=1}^K )$. Let $E_k=\{(j,j^\prime): j \leq j^\prime, \omega^{k}_{j,j^\prime} \neq 0 \}$ be the set of indices of nonzero off-diagonal elements in $\Omega_k$, and $E_0=\{(j,j^\prime): j \leq j^\prime, \omega^{0}_{j,j^\prime} \neq 0 \}$ be the set of indices of nonzero off-diagonal elements in $\Omega_0$. Let $q_k=|E_k|$ and $q_0=|E_0|$ be the cardinalities of $E_k$ and $E_0$, with $q_k,q_0 \leq q$. The main idea of the proof follows Rothman et al. (2008) and Lam and Fan
(2009). Let 
\bse
Q(\Omega ) = \sum_{k=1}^K \left\{ -\log\det(\Omega_k) + \mathtt{tr}(S_k\Omega_k) \right\} + P(\Omega_0,\left\{ \Omega_k \right\})
\ese
with the penalty term $P(\Omega)$ specified as in (\ref{eq:penalty}), and
\bse
G(\Delta) &=&  Q(\Omega+\Delta) -  Q(\Omega) \\
	&=& \sum_{k=1}^K \mathtt{tr} \left[ \left\{ S_k+\lambda_2(\Omega_0+\Delta_0)^{-1} \right\}  (\Omega_0+\Delta_0) - \left\{ S_k + \lambda_2\Omega_0^{-1}\right\} \Omega_0 \right] \\
	& & -\sum_{k=1}^K(1+\lambda_2) \left\{ \log\det(\Omega_k+\Delta_k)-\log\det(\Omega_k)\right\} \\
	& & + K\lambda_2 \left\{ \log\det(\Omega_0+\Delta_0)-\log\det(\Omega_0)\right\} \\
	& & + \sum_{k=1}^K \lambda_1 \left\{ |\Omega_k+\Delta_k|-|\Omega_k| \right\} 
		+\lambda_3 \left\{ |\Omega_0+\Delta_0|-|\Omega_0| \right\}.
\ese
We want to show that, for $\alpha_n=(q\log p/n)^{1/2}$ and $\beta_n=(p\log p/n)^{1/2}$, and for the set $\mathcal{A}$ defined as $\mathcal{A}=\{\Delta: \sum_{k=1}^K||\Delta_k||^2_F=C_1^2\alpha_n^2+C_2^2\beta_n^2, \text{ and } ||\Delta_0||^2_F=C_3^2\alpha_n^2+C_4^2\beta_n^2 \}$, 
\bse
P \left( \inf_{\Delta)\in \mathcal{A}} G(\Delta) > 0 \right) 
	\rightarrow 1
\ese
for sufficiently large constants $C_1, C_2, C_3,$ and $C_4$. This implies that there is a local minimizer in $\{ \Omega+\Delta: \sum_{k=1}^K||\Delta_k||^2_F=C_1^2\alpha_n^2+C_2^2\beta_n^2, \text{ and } ||\Delta_0||^2_F=C_3^2\alpha_n^2+C_4^2\beta_n^2 \}$ such that $\sum_{k=1}^K ||\widehat{\Omega}_k-\Omega_k||_F= O_P(\sqrt{(q+p)logp/n})$ and $||\widehat{\Omega}_0-\Omega_0||_F= O_P(\sqrt{(q+p)logp/n})$ for sufficiently large $n$.

Using Taylor's expansion with the integral form of the remainders, we can write 
\bse
G(\Delta) = I_1+I_2+I_3+I_4+I_5,
\ese
where
\bse
I_1 & = & \sum_{k=1}^K \mathtt{tr}\{ (S_k-\Sigma_k) \Delta_k \} 
	+ K\lambda_2 \left[ - \mathtt{tr} \left\{ (\bar{\Omega}_k - \Omega_0)\Sigma_0 \Delta_0 \Sigma_0 \right\}   \right] \\
	&& + \lambda_2 \sum_{k=1}^K \mathrm{tr}\left\{ ((\Omega_0+\Delta_0)^{-1}-\Sigma_k) \Delta_k \right\} \\
I_2 & = & (1+\lambda_2) \sum_{k=1}^K vec(\Delta_k)^T \left\{ \int^1_0 g(v,\Omega_k^v)(1-v)dv \right\} vec(\Delta_k)  \\
I_3 & = & K\lambda_2 \left[ vec(\Delta_0)^T \left\{ \int^1_0 f(v,\Omega_0^v)(1-v)dv \right\} vec(\Delta_0) \right] \\
I_4 & = & \lambda_1 \sum_{k=1}^K \sum_{(j,j^\prime) \in E_k} (|\omega^k_{jj^\prime}+\delta^k_{jj^\prime}| - |\omega^k_{jj^\prime}| ) 
		+ \lambda_3 \sum_{(j,j^\prime) \in E_0} (|\omega^0_{jj^\prime}+\delta^0_{jj^\prime}| - |\omega^0_{jj^\prime}| )\\
I_5 & = & \lambda_1 \sum_{k=1}^K \sum_{(j,j^\prime) \in E_k^C} |\delta^k_{jj^\prime}| 
		+ \lambda_3 \sum_{(j,j^\prime) \in E_0^C} |\delta^0_{jj^\prime}| ,
\ese
with $\bar{\Omega}_k=\sum_{k=1}^K\Omega_k/K \equiv\Omega_0$, $\Omega_k^v=\Omega_k+v \Delta_k$, $\Omega_0^v=\Omega_0+v \Delta_0$, $g(v,\Omega_k^v)=(\Omega_k^v)^{-1} \otimes (\Omega_k^v)^{-1}$, and $f(v,\Omega_0^v)=(\Omega_0^v)^{-1} \otimes (\Omega_0^v)^{-1}\bar{\Omega}_k(\Omega_0^v)^{-1} + (\Omega_0^v)^{-1}\bar{\Omega}_k(\Omega_0^v)^{-1} \otimes (\Omega_0^v)^{-1} - (\Omega_0^v)^{-1} \otimes (\Omega_0^v)^{-1}$.

By condition (A1) and $\lambda_2=O(\sqrt{\log p/n})=o(1)$, we have
\bse
I_2 &=& (1+\lambda_2) \sum_{k=1}^K vec(\Delta_k)^T \left\{ \int^1_0 (\Omega_k^v)^{-1} \otimes (\Omega_k^v)^{-1}(1-v)dv \right\} vec(\Delta_k)  \\	
	&\geq & (1+\lambda_2) \sum_{k=1}^K   \frac{1}{2} ||\Delta_k||_F^2 \min_{0 \leq v \leq 1} \psi_{\max}^{-2}(\Omega^v_k)  \\
	& \geq & (1+\lambda_2) (||\Omega_k||+||\Delta_k||)^{-2}/2 \cdot \sum_{k=1}^K  ||\Delta_k||_F^2   \\ 
	& \geq &(1+o(1)) \cdot (\tau_2+o(1))^{-2}/2 \cdot \sum_{k=1}^K  ||\Delta_k||_F^2  \\
	& \geq & (\tau_2^{-2}/2 +o(1))  \cdot (C_1^2\alpha_n^2+C_2^2\beta_n^2).
\ese

By Neumann series expansion, we have
$$ 
(\Omega^v_0)^{-1} = \Sigma_0(I+v\Delta_0)^{-1}=\Sigma_0(I-v\Delta_0\Omega_0 + o(1)),
$$
which means $(\Omega^v_0)^{-1}=\Sigma_0+O_P(\alpha_n+\beta_n)$, and $||(\Omega^v_0)^{-1}||=\tau_1^{-1}+O_P(\alpha_n+\beta_n)$. With $\bar{\Omega}_k=\Omega_0$,
$$
\bar{\Omega}_k(\Omega_0^v)^{-1} = I-v\Delta_0\Omega_0 + o(1) = I + o(1).
$$
Combining these, we have $f(v,\Omega_0^v)= \Sigma_0 \otimes \Sigma_0 + O_P(\alpha_n+\beta_n)$, and therefore
\bse
I_3 &=& K \lambda_2 \left[ vec(\Delta_0)^T \left\{ \int^1_0 \Sigma_0 \otimes \Sigma_0(1+o_P(1))(1-v)dv \right\} vec(\Delta_0) \right] \\	
	&\geq & K \lambda_2 \cdot \frac{1}{2} ||\Delta_0||_F^2  \cdot \psi_{\min}(\Sigma_0 \otimes \Sigma_0) (1+o_P(1)) \\
	& \geq & K \lambda_2  \cdot \frac{1}{2} ||\Delta_0||_F^2 \cdot \psi_{\min}(\Sigma_0 \otimes \Sigma_0)  (1+o_P(1)) \\
	& \geq & \lambda_2 \cdot K\tau_2^{-2}(1+o_P(1))/2 \cdot ||\Delta_0||_F^2 \\
	& \geq & \lambda_2 \cdot (K\tau_2^{-2}/2 +o_P(1)) \cdot (C_3^2\alpha_n^2+C_4^2\beta_n^2) .
\ese

Now consider $I_1$. Using again the Neumann expansion for $(\Omega_0+\Delta_0)^{-1}$ and with $\bar{\Omega}_k=\Omega_0$, we have 
\bse
I_1 &=& \sum_{k=1}^K \mathtt{tr}\{ (S_k-\Sigma_k) \Delta_k \} 
	+ \lambda_2 \sum_{k=1}^K \mathrm{tr}\left\{ (\Sigma_0(I-\Delta_0\Sigma_0+o(1))-\Sigma_k) \Delta_k \right\} \\
	&=& \sum_{k=1}^K \mathtt{tr}\{ (S_k-\Sigma_k) \Delta_k \} 
	+ \lambda_2\sum_{k=1}^K \mathrm{tr}\left\{\Sigma_0(\Omega_k-\Omega_0)\Sigma_k\Delta_k \right\} -\lambda_2 \sum_{k=1}^K \mathrm{tr}\left\{ \Sigma_0\Delta_0\Sigma_0\Delta_k \right\} (1+o(1)).
\ese
It is clear that $|I_1| \leq L_1 + L_2$, where
\bse
L_1 &=& \sum_{k=1}^K \left|\sum_{(j,j^\prime) \in E_k} (S_k-\Sigma_k)_{jj^\prime}\delta^k_{jj^\prime} \right| + \lambda_2 \sum_{k=1}^K \left| \sum_{(j,j^\prime) \in E_k} (\Sigma_0(\Omega_k-\Omega_0)\Sigma_k)_{jj^\prime}\delta^k_{jj^\prime} \right|  \\
	&&  + \lambda_2 \sum_{k=1}^K \left| \sum_{(j,j^\prime) \in E_k} (\Sigma_0\Delta_0\Sigma_0)_{jj^\prime}\delta^k_{jj^\prime} \right| (1+o(1))   \; ,\\
L_2 &=& \sum_{k=1}^K \left|\sum_{(j,j^\prime) \in E_k^c} (S_k-\Sigma_k)_{jj^\prime}\delta^k_{jj^\prime} \right| + \lambda_2 \sum_{k=1}^K \left| \sum_{(j,j^\prime) \in E_k^c} (\Sigma_0(\Omega_k-\Omega_0)\Sigma_k)_{jj^\prime}\delta^k_{jj^\prime} \right| \\
	&&  + \lambda_2 \sum_{k=1}^K \left| \sum_{(j,j^\prime) \in E_k^c} (\Sigma_0\Delta_0\Sigma_0)_{jj^\prime}\delta^k_{jj^\prime} \right| (1+o(1)) \; .
\ese
By condition (A2) and $||\Delta_0|| \leq ||\Delta_0||_F=o(1)$, and using Lemmas 1 \& 2 and , we have
\bse
L_1 & \leq & \sum_{k=1}^K (q_k+p)^{1/2} \max_{j,j^\prime} |(S_k-\Sigma_k)_{jj^\prime}|\cdot ||\Delta_k||_F \\
	&& + \lambda_2 \sum_{k=1}^K (q_k+p)^{1/2} \max_{j,j^\prime} |(\Sigma_0(\Omega_k-\Omega_0)\Sigma_k)_{jj^\prime}|\cdot ||\Delta_k||_F \\
	&& + \lambda_2 \sum_{k=1}^K (q_k+p)^{1/2} \max_{j,j^\prime} |(\Sigma_0\Delta_0\Sigma_0)_{jj^\prime}| \cdot ||\Delta_k||_F \cdot (1+o(1)) \\
	& \leq & \sum_{k=1}^K O_P(\sqrt{(q_k+p)\log p/n}) \cdot ||\Delta_k||_F \\
	&& + \sum_{k=1}^K O(\sqrt{(q_k+p)\log p/n}) \cdot \epsilon\tau_1^{-2}  \cdot ||\Delta_k||_F \\
	&& + \sum_{k=1}^K O(\sqrt{(q_k+p)\log p/n}) \cdot \tau_1^{-2}||\Delta_0||  \cdot ||\Delta_k||_F \\
	& \leq & O_P(\alpha_n+\beta_n) \cdot (1+\epsilon \tau_1^{-2}+o(1)) \cdot \sum_{k=1}^K ||\Delta_k||_F \\
	&=& O_P(C_1 \alpha_n^2 + C_2 \beta_n^2).
\ese
$L_1$ is thus dominated by $I_2$ when $C_1$ and $C_2$ are sufficiently large.

Now consider $I_3$. By the triangular inequality, we have $|I_4| \leq H_1+H_2$, where
\bse
H_1 &=& \sum_{k=1}^K \lambda_1 \sum_{(j,j^\prime) \in E_k} |\delta^k_{jj^\prime}| \\
  & \leq & \sum_{k=1}^K \lambda_1 (q_k+p)^{1/2} ||\Delta_k||_F  \\
  & \leq & O(\alpha_n+\beta_n) \cdot \sum_{k=1}^K ||\Delta_k||_F \\
  & = & O_P(C_1 \alpha_n^2 + C_2 \beta_n^2),
\ese
since $\lambda_1 \asymp \sqrt{\log p/n}$. Thus, $H_1$ is dominated by $I_2$. Similarly, 
\bse
H_2 &=& \lambda_3 \sum_{(j,j^\prime) \in E_0} |\delta^0_{jj^\prime}|  \\
  & \leq & \lambda_3 (q_0+p)^{1/2}||\Delta_0||_F \\
  & \leq & \lambda_2 \cdot O(\alpha_n+\beta_n) \cdot ||\Delta_0||_F  \\
  & = & \lambda_2 \cdot O_P(C_3 \alpha_n^2 + C_4 \beta_n^2).
\ese
Since $\lambda_3/\lambda_2 = O(\sqrt{\log p/n})$, $H_2$ is dominated by $I_3$.

Now with $L_1$ and $H_1$ dominated by $I_2$ and $H_2$ dominated by $I_3$, the proof completes if we can show $I_5-L_2 \geq 0$. 
\bse
I_4 - L_2 & \geq & \sum_{k=1}^K \sum_{(j,j^\prime) \in E_k^c} \Big\{ \lambda_1 -|(S_k-\Sigma_k)_{jj^\prime}| - \lambda_2 |(\Sigma_0(\Omega_k-\Omega_0)\Sigma_k)_{jj^\prime} |\\
	&& - \lambda_2 |(\Sigma_0\Delta_0\Sigma_0)_{jj^\prime}|(1+o(1)) \Big\} |\delta^k_{jj^\prime}| 
	\;\; + \sum_{(j,j^\prime) \in E_k^c} \lambda_3  |\delta^0_{jj^\prime}|  \\
	& \geq & \sum_{k=1}^K \sum_{(j,j^\prime) \in E_k^c} \Big\{ \lambda_1 - |(S_k-\Sigma_k)_{jj^\prime}|
	- \lambda_2 (\epsilon \tau_1^{-2} + o(1)) \Big\} |\delta^k_{jj^\prime}| 
	\;\; + \sum_{(j,j^\prime) \in E_k^c} \lambda_3  |\delta^0_{jj^\prime}|.
\ese
Since $\max_{j\neq j^\prime}|(S_k-\Sigma_k)_{jj^\prime}|=O_P(\sqrt{\log p/n})$, and $\lambda_2=O(\sqrt{\log p/n})$, we can find a positive $W_1=O_P(1)$ and $W_2=O(1)$ such that 
$$
\max_{j\neq j^\prime}|(S_k-\Sigma_k)_{jj^\prime}|=W_1 \sqrt{\log p/n} , \text{ and }
\lambda_2 = W_2 \sqrt{\log p/n}.
$$
Then we can find $\lambda_1 = W_3 \sqrt{\log p/n}$ with $W_3 > W_1+\epsilon \tau_1^{-2}W_2$, so that $I_4-L_2 \geq 0$. This completes the proof of the theorem.  \hfill $\square$

\vspace{.2in}
{\it Proof of Theorem 2.} For $\widehat{\Omega}=(\widehat{\Omega}_0,\{\widehat{\Omega}_k\}_{k=1}^K)$ a minimizer of the objective function $Q$, the derivative for $Q$ with respect to $\omega^k_{jj^\prime}$ for $(j,j^\prime) \in E^c_k$ and $\omega^0_{jj^\prime}$ for $(j,j^\prime) \in E^c_0$ evaluated at $\widehat{\Omega}$ are, respectively,
\bse
\left. \frac{\partial Q}{\partial \omega^k_{jj^\prime}} \right|_{\widehat{\Omega}} 
	&=& 2 \left\{ s^k_{jj^\prime}+\lambda_2\hat{\sigma}^0_{jj^\prime} - (1+\lambda_2)\hat{\sigma}^k_{jj^\prime} + \lambda_1 \text{sgn}(\hat{\omega}^k_{jj^\prime}) \right\}  \; , \\
\left. \frac{\partial Q}{\partial \omega^0_{jj^\prime}} \right|_{\widehat{\Omega}} 
	&=& 2 \lambda_2 \left\{ -\sum_{k=1}^K \left( \widehat{\Sigma}_0 \widehat{\Omega}_k \widehat{\Sigma}_0 \right)_{jj^\prime} + K \hat{\sigma}^0_{jj^\prime} + \frac{\lambda_3}{\lambda_2} \text{sgn}(\hat{\omega}^0_{jj^\prime}) \right\}  \; . 
\ese
If we can show that the sign of $\partial Q/\partial \omega^k_{jj^\prime}$ evaluated at $\widehat{\Omega}$ depends on $\text{sgn}(\hat{\omega}^k_{jj^\prime})$ only with probability tending to 1, the optimum will be at 0, so that $\hat{\omega}^k_{jj^\prime}=0$ for all ${jj^\prime} \in E^c_k$ with probability tending to 1. Similarly, to prove $\hat{\omega}^0_{jj^\prime}=0$ for all ${jj^\prime} \in E^c_0$ with probability tending to 1, it suffices to show that the sign of $\partial Q/\partial \omega^0_{jj^\prime}$ evaluated at $\widehat{\Omega}$ has the same sign as $\hat{\omega}^0_{jj^\prime}$  with probability tending to 1. 

First, for sparsity of $\widehat{\Omega}_k$ ($k=1,\ldots,K$),
\bse
s^k_{jj^\prime}+\lambda_2\hat{\sigma}^0_{jj^\prime} - (1+\lambda_2)\hat{\sigma}^k_{jj^\prime} 
	&=& (s^k_{jj^\prime} - \hat{\sigma}^k_{jj^\prime} ) + \lambda_2 (\hat{\sigma}^0_{jj^\prime} - \hat{\sigma}^k_{jj^\prime}) \\
 	&=& I_1 + I_2+I_3+I_4,
\ese
where 
\bse
I_1 = s^k_{jj^\prime} - \sigma^k_{jj^\prime}, 
\quad I_2 = (1+\lambda_2) (\sigma^k_{jj^\prime} - \hat{\sigma}^k_{jj^\prime}) , 
\quad I_3 = \lambda_2 (\hat{\sigma}^0_{jj^\prime} - \sigma^0_{jj^\prime}),
\quad I_4 = \lambda_2 (\sigma^0_{jj^\prime}-\sigma^k_{jj^\prime}).
\ese

By Lemma 2, $\max_{j,j^\prime}|I_1|=O_P(\sqrt{\log p/n})$. By Lemma 1, 
\bse
|I_2| \leq (1+\lambda_2) ||\widehat{\Sigma}_k-\Sigma_k|| \leq (1+\lambda_2) ||\widehat{\Sigma}_k|| \cdot ||\widehat{\Omega}_k-\Omega_k|| \cdot ||\Sigma_k|| = O(||\widehat{\Omega}_k-\Omega_k||) = O(\sqrt{\eta_n}),
\ese
since $\lambda_2=O(\sqrt{\log p/n})=o(1)$, $||\Sigma_k||=O(1)$ by condition (A1), and 
\bse
||\widehat{\Sigma}_k|| = \psi^{-1}_{\min}(\widehat{\Omega}_k) \leq (\psi_{\min}(\Omega_k)+\psi_{\min}(\widehat{\Omega}_k-\Omega_k))^{-1} = (O(1)+o(1))^{-1}=O(1).
\ese
Similarly,
\bse
|I_3| \leq \lambda_3||\widehat{\Sigma}_0-\Sigma_0|| \leq \lambda_2 ||\widehat{\Sigma}_0|| \cdot ||\widehat{\Omega}_0-\Omega_0|| \cdot ||\Sigma_0|| = \lambda_2 \cdot O(||\widehat{\Omega}_0-\Omega_0||) = o(\sqrt{\eta_n}),
\ese
and
\bse
|I_4| \leq \lambda_2 ||\Sigma_k-\Sigma_0|| \leq \lambda_2||\Sigma_k||\cdot||\Omega_k-\Omega_0|| \cdot ||\Sigma_0|| < \epsilon \lambda_2 \cdot O(1) = O(\sqrt{\log p/n}),
\ese
since $||\Omega_k-\Omega_0|| < \epsilon$.

Combining all these results yields that
\bse
\max_{jj^\prime}|s^k_{jj^\prime}+\lambda_2\hat{\sigma}^0_{jj^\prime} - (1+\lambda_2)\hat{\sigma}^k_{jj^\prime}| = O_P(\sqrt{\log p/n}+\sqrt{\eta_n}).
\ese
Therefore, we need to have $\sqrt{\log p/n}+\sqrt{\eta_n} = O(\lambda_1)$ in order to have the sign of $(\partial Q/\partial \omega^k_{jj^\prime})|_{\widehat{\Omega}}$ depends on $\text{sgn}(\hat{\omega}^k_{jj^\prime})$ with probability tending to 1.

Now, for sparsity of $\widehat{\Omega}_0$, 
\bse
\left| -\sum_{k=1}^K \left( \widehat{\Sigma}_0 \widehat{\Omega}_k \widehat{\Sigma}_0 \right)_{jj^\prime} + K \hat{\sigma}^0_{jj^\prime} \right| 
	&=& \left| \left(-\sum_{k=1}^K \left(\widehat{\Sigma}_0 \widehat{\Omega}_k \widehat{\Sigma}_0 \right)+ K\widehat{\Sigma}_0 \right)_{jj^\prime} \right| \\
	&=& \left| \left(\sum_{k=1}^K L_{1k}+ K L_2 \right)_{jj^\prime} \right|,
\ese
with
\bse
L_{1k} = - \widehat{\Sigma}_0 \left( \widehat{\Omega}_k -\Omega_k \right) \widehat{\Sigma}_0 \; , 
	\quad L_2 = \widehat{\Sigma}_0 \left( \widehat{\Omega}_0 -\Omega_0 \right) \widehat{\Sigma}_0 ,
\ese
where we used $\Omega_0=\sum_{k=1}^K\Omega_k/K$ by condition (A2).

Since 
\bse
||\widehat{\Sigma}_0|| = \psi^{-1}_{\min}(\widehat{\Omega}_0) \leq (\psi_{\min}(\Omega_0)+\psi_{\min}(\widehat{\Omega}_0-\Omega_0))^{-1} = (O(1)+o(1))^{-1}=O(1),
\ese
we have
\bse
&& \max_{jj^\prime}|(L_{1k})_{jj^\prime}| \leq ||\widehat{\Sigma}_0|| \cdot || \widehat{\Omega}_k -\Omega_k || \cdot || \widehat{\Sigma}_0 || = O(|| \widehat{\Omega}_k -\Omega_k ||) = O(\sqrt{\eta_n}) \; ,\\
\text{and } 
&& \max_{jj^\prime}|(L_2)_{jj^\prime}| \leq ||\widehat{\Sigma}_0|| \cdot || \widehat{\Omega}_0 -\Omega_0 || \cdot || \widehat{\Sigma}_0 || = O(|| \widehat{\Omega}_0 -\Omega_0 ||) = O(\sqrt{\eta_n}) \; ,
\ese
and consequently, 
\bse
\left| -\sum_{k=1}^K \left( \widehat{\Sigma}_0 \widehat{\Omega}_k \widehat{\Sigma}_0 \right)_{jj^\prime} + K \hat{\sigma}^0_{jj^\prime} \right| = O(\sqrt{\eta_n}).
\ese
Therefore, we need to have $\sqrt{\eta_n} = O(\lambda_3/\lambda_2)$ in order to have the sign of $(\partial Q/\partial \omega^0_{jj^\prime})|_{\widehat{\Omega}}$ depends on $\text{sgn}(\hat{\omega}^0_{jj^\prime})$ with probability tending to 1.  \hfill $\square$


\end{document}